Article

# THE ANTHROPIC PRINCIPLE IS TESTABLE AND APPEARS WEAK

STUART KAUFFMAN
Institute for Systems Biology
Correspondence: stukauffman@gmail.com

MAY 22, 2022

ABSTRACT

The Anthropic Principle has been with us since the 1970s. This Principle is advanced to account for the "fine tuning" of the 25 constants of the Standard Model of Particle Physics. Were these constants very different, life could not exist. The Anthropic Principle conditions on the existence of life and concludes that the value of the 25 constants must be within a range that allows life. The most common further step is to postulate the existence of a vast multiverse with vastly many combinations of the values of the 25 constants. By conditioning on our own life, we must be in a universe whose values allow life.

The Anthropic Principle is commonly held to be untestable because we cannot be in contact with other universes. I aim here to show the Anthropic Principle *is* testable and that its explanatory power is weak: The Principle seems to make testably false predictions about planet Earth and the life on it. The Anthropic Principle seems unable to predict the existence of 98 stable atoms, when only 19 small atoms are needed for life.

Key Words: Anthropic Principle, Multiverse, 98 stable atoms; 19 atoms required for life; Incapacity of the Anthropic Principle to explain why there are 98 stable atoms.

## INTRODUCTION

### THE FUNDAMENTAL ISSUE: WHY THESE LAWS AND CONSTANTS?

The current foundational laws of physics are The Standard Model of Particle Physics, and General Relativity. The two are not united. Both are confirmed to thirteen decimal places, one part in ten trillion.

These laws have a number of numerical constants that are not derived from theory, so are in that sense "arbitrary" values utilized in the equations to fit the data. The Standard Model of particle physics has 25 parameters. General Relativity has 2 parameters.

Given any set of laws, Newton's, or our current set, we can ask a fundamental question: Why these laws and constants?

The first class of answers to this question is that we cannot even ask the question. The laws and constants just *are* what they *are*. Rather like God, the laws, in some sense, just exist outside the universe and "govern" or "describe" the universe.

A second answer to the question, "Why these laws and constants" is that this is due to God.

## THE ANTHROPIC PRINCIPLE AND THE MULTIVERSE

Five decades ago, it became clear that the 25 constants of the standard model must be "tuned" very finely in our universe. Small variations in the constants predicted radically different universes. For most such values of the constants, life could not exist. We are living and exist, hence, why the fine tuning?

### The Anthropic Principle

A dominant effort to answer the question of the fine tuning of the constants is based on the fact that life exists, and intelligent observers exist who can observe the constants of nature. If we condition on the fact that intelligent life exists, then the universe must have the values of the constants that allow life or intelligent life to exist. This is the "Anthropic Principle". This Principle comes in a few versions, two are below.

**Weak anthropic principle (WAP)** (Carter (1)): "[W]e must be prepared to take account of the fact that our location in the universe is *necessarily* privileged to the extent of being compatible with our existence as observers." Note that for Carter, "location" refers to our location in time as well as space, (2).

**Strong anthropic principle (SAP)** (Carter (3)): "[T]he universe (and hence the fundamental parameters on which it depends) must be such as to admit the creation of observers within it at some stage. To paraphrase Descartes, *cogito ergo mundus talis est*."

The Latin tag ("I think, therefore the world is such [as it is]") makes it clear that "must" indicates a deduction from the fact of our existence; the statement is thus a truism, (2).

### The Multiverse

The most familiar way to realize the Anthropic Principle is to posit a vast ensemble of universes, a multiverse, *ME*, with all possible combinations of values of the 25 constants. Only a tiny subset of these, *LE*, have values of the constants that are compatible with life or intelligent life. Since we are alive, or also intelligent, we necessarily exist in a universe in the LE sub-ensemble of universes that allow life, or intelligent life, (1,2,3,4).

Thus, due to this Anthropic Principle, we can expect to find ourselves, because alive and intelligent, in a universe whose constants permit such life. The fine tuning of the constants is explained.

A long history of critique of the Anthropic Principle exists. Among many other issues, the greatest concern, since we can be in no contact with the other universes, is that the Anthropic Principle has been held to be untestable, (5,6,7).

Testing the Anthropic Principle

While we cannot be in contact with other universes, there are two related ways to test the explanatory power of the Anthropic Principle. Both are critical. The aim of the Anthropic Principle is to explain our specific values of the 25 constants.

i. If the Life Ensemble, LE, is a *large* sub-volume of the total ensemble of universes, ME, then our membership in LE does not explain our specific values of the 25 constants. Thus, as our proposal for the requirements of life become broader, the volume of LE within ME increases, and the explanatory power of the Anthropic Principle fades.

ii. For the Anthropic Principle to be explanatory of our value of the 25 constants, those values must be *typical* of the range of values of the constants of LE, (8).

iii. The consequences of the issues above are that the explanatory power of the Anthropic Principle is maximized if the ratio of LE/ME is minimized, while the typicality of our values of the constants in LE is sustained.

iv. Based on these issues, I aim to show that that the explanatory power of the Anthropic Principle is weak.

How many stable atoms are allowed by the Standard Model? In their book, "The Grand Design", (9), Hawking and Mlodinow point out that the masses of the quarks are tuned to maximize the number of different stable atoms. 98 different stable atoms exist naturally. The potential number of stable atoms is debated, (10).

In considering the volume of LE, as constants are tuned such that number of stable atoms could increase, say to 300, and many or most of these are also presumed requisite for Life, the ratio of LE/ME increases. Therefore, the power of the Anthropic Principle to explain the *specific* values of our constants decreases. The explanatory power of the Anthropic principle is increased by *minimizing* LE.

**The atoms known to be requisite for life on Earth.**

Our universe does have 98 stable atoms. Only a small subset of these, a total of 19 different atoms, is needed for life and a rocky planet with a crust to host that life. These atoms are all small. The maximum atomic number is 30. Table 1 gives the set of atoms needed for earth life, (11,12).

TABLE 1

ATOMS IN BIOLOGY

| Atom | Atomic Number |
| --- | --- |
| H | 1 |
| C | 6 |
| N | 7 |
| O | 8 |
| Na | 11 |
| Mg | 12 |
| Al | 13 |
| Si | 14 |
| Mn | 14 |
| P | 15 |
| S | 16 |
| Cl | 17 |
| K | 19 |
| Mn | 25 |
| Fe | 26 |
| Co | 27 |
| Ni | 28 |
| Cu | 29 |
| Zn | 30 |

Let us call our particular universe to be one in a restricted sub-ensemble, 98E *within* the life ensemble, LE, having the 98 stable atoms. Our ensemble, 98E, is a subset of the Life Ensemble, LE.

We can now test the Anthropic Principle. *If our sub-ensemble, 98E is a tiny sub-volume of the life ensemble, LE, and if the members of 98E are not randomly located in LE, then our universe is not a typical member* of the life ensemble, LE.

If our universe is, indeed, not a typical member of the Life Ensemble, LE, then an Anthropic Principle that conditions only of the fact that we are alive would offer no explanation for the fact that our universe has 98 stable atoms. We would have no explanation for our values of the constants.

A proper test of the Anthropic principle requires assessing the measure of the sub-volume, 98E, of the Life Ensemble, LE, over all values of the 25 constants. If 98E/LE is small, say 0.001, or 0.000001, then the Anthropic Principle is correspondingly weakened.

The conclusion that the Anthropic Principle is weakened is further supported if 98E is not typical of LE, (8), i.e. *non-randomly distributed* in LE.

Because the formation of larger atomic nuclei in nucleosynthesis depends upon the prior formation of smaller nuclei from which the larger nuclei are built in stars and supernovae, (13), it seems likely that 98E is not randomly located in LE.

In so far as additional constraints on the constants are required for synthesis of atoms of atomic numbers greater than 30, it is also true the 98E ensemble is a correspondingly small sub-ensemble of the Life Ensemble, LE.

It may or may not be feasible to carry out such a calculation in detail of the volumes of LE and 98E or to estimate a non- random distribution of 98E in LE. And there may be doubts about exactly how many types of atoms are needed for life. However, in so far as it now seems likely that [98E/ LE] is very small and 98E is not typical of LE, the explanatory power of the Anthropic Principle seems weakened.

CONCLUSION

I reach the tentative conclusion, pending better evidence, that the explanatory power of the Anthropic Principle is weak. In so far as this is true, we have no clear grounds to account for the values of constants that we observe.

More, insofar as the multiverse is held as a hypothesis to explain the values of the constants, this grounds to support the existence of a multiverse to explain the constants is also weakened.

The Author has the right to publish this material and has no conflicting interests.

This work was not supported by grants.

Acknowledgements

The Author is glad to thank George Ellis for several fruitful discussions and very useful critical reviews.